\documentclass[apj,iop,twocolumn]{emulateapj}
\usepackage{amsmath}
\usepackage{float}
\usepackage{nicefrac}
\usepackage{rotating}
\usepackage{rotfloat}
\usepackage[none]{hyphenat}
\usepackage{enumitem}
\usepackage{afterpage}

\def\aj{AJ}

\def\apj{ApJ}

\def\apss{Ap\&SS}
\def\aap{A\&A}

\def\aaps{A\&AS}

\def\mnras{MNRAS}

\def\pasa{Publ.~Astron.~Soc.~Australia}
\def\pasp{PASP}

\def\rmxaa{RMxAA}

\def\nar{New~Astro.~Rev.}

\def\arcmin{\hbox{$^\prime$}}
\def\arcsec{\hbox{$^{\prime\prime}$}}

\newcommand{\eld}{${\it N}_{\rm e}$}

\newcommand{\elt}{${\it T}_{\rm e}$}

\newcommand{\foiii}{[O\,{\sc iii}]}

\newcommand{\fsii}{[S\,{\sc ii}]}
\newcommand{\fsiii}{[S\,{\sc iii}]}

\newcommand{\fnii}{[N\,{\sc ii}]}

\newcommand{\fariv}{[Ar\,{\sc iv}]}

\newcommand{\ffeiii}{[Fe\,{\sc iii}]}

\newcommand{\fkiv}{[K\,{\sc iv}]}

\newcommand{\nii}{N\,{\sc ii}}
\newcommand{\niii}{N\,{\sc iii}}

\newcommand{\oii}{O\,{\sc ii}}
\newcommand{\cii}{C\,{\sc ii}}

\newcommand{\hi}{H\,{\sc i}}

\newcommand{\hei}{He\,{\sc i}}
\newcommand{\heii}{He\,{\sc ii}}

\newcommand{\hb}{H$\beta$}

\shorttitle{FLIERs of M2-42}
\shortauthors{Danehkar, Parker, \& Steffen}

\slugcomment{AJ; submitted 2015 March 18; accepted 2015 December 18}

\begin{document}

\title{Fast, low-ionization emission regions of the planetary nebula M2-42}

\author{A. Danehkar\altaffilmark{1,2}, Q. A. Parker\altaffilmark{1,3,4}, and W.~Steffen\altaffilmark{5}}
\email{ashkbiz.danehkar@cfa.harvard.edu}

\altaffiltext{1}{Department of Physics and Astronomy, Macquarie University, Sydney, NSW 2109, Australia}
\altaffiltext{2}{Harvard-Smithsonian Center for Astrophysics, 60 Garden Street, Cambridge, MA 02138, USA}
\altaffiltext{3}{Australian Astronomical Observatory, P.O. Box 915, North Ryde, NSW 1670, Australia}
\altaffiltext{4}{Department of Physics, The University of Hong Kong, Pokfulam Road, Hong Kong, China}
\altaffiltext{5}{Instituto de Astronom\'{i}a, Universidad Nacional Aut\'{o}noma de M\'{e}xico, C.P.22860, Ensenada, Mexico}

\begin{abstract}Spatially resolved observations of the planetary nebula M2-42 (PN\,G008.2$-$04.8) obtained with the Wide Field Spectrograph on the Australian National University 2.3 m telescope have revealed the remarkable features of bipolar collimated jets emerging from its main structure. Velocity-resolved channel maps  derived from the [N\,{\sc ii}] $\lambda$6584 emission line disentangle different morphological components of the nebula. This information is used to develop a three-dimensional morpho-kinematic model, which consists of an equatorial dense torus and a pair of asymmetric bipolar outflows. The expansion velocity of about 20 km\,s$^{-1}$ is measured from the spectrum integrated over the main shell. However, the deprojected velocities of the jets are found to be in the range of $80$--160 km\,s$^{-1}$ with respect to the nebular center. It is found that the mean density of the collimated outflows, $595 \pm 125$\,cm$^{-3}$, is five times lower than that of the main shell, $3150$\,cm$^{-3}$, whereas their singly ionized nitrogen and sulfur abundances are about three times higher than those determined from the dense shell. The results indicate that the features of the collimated jets are typical of fast, low-ionization emission regions. 
\end{abstract}


\section{Introduction}
\label{m2_42:sec:introduction}

\begin{figure}
\begin{center}
\includegraphics[width=2.4in]{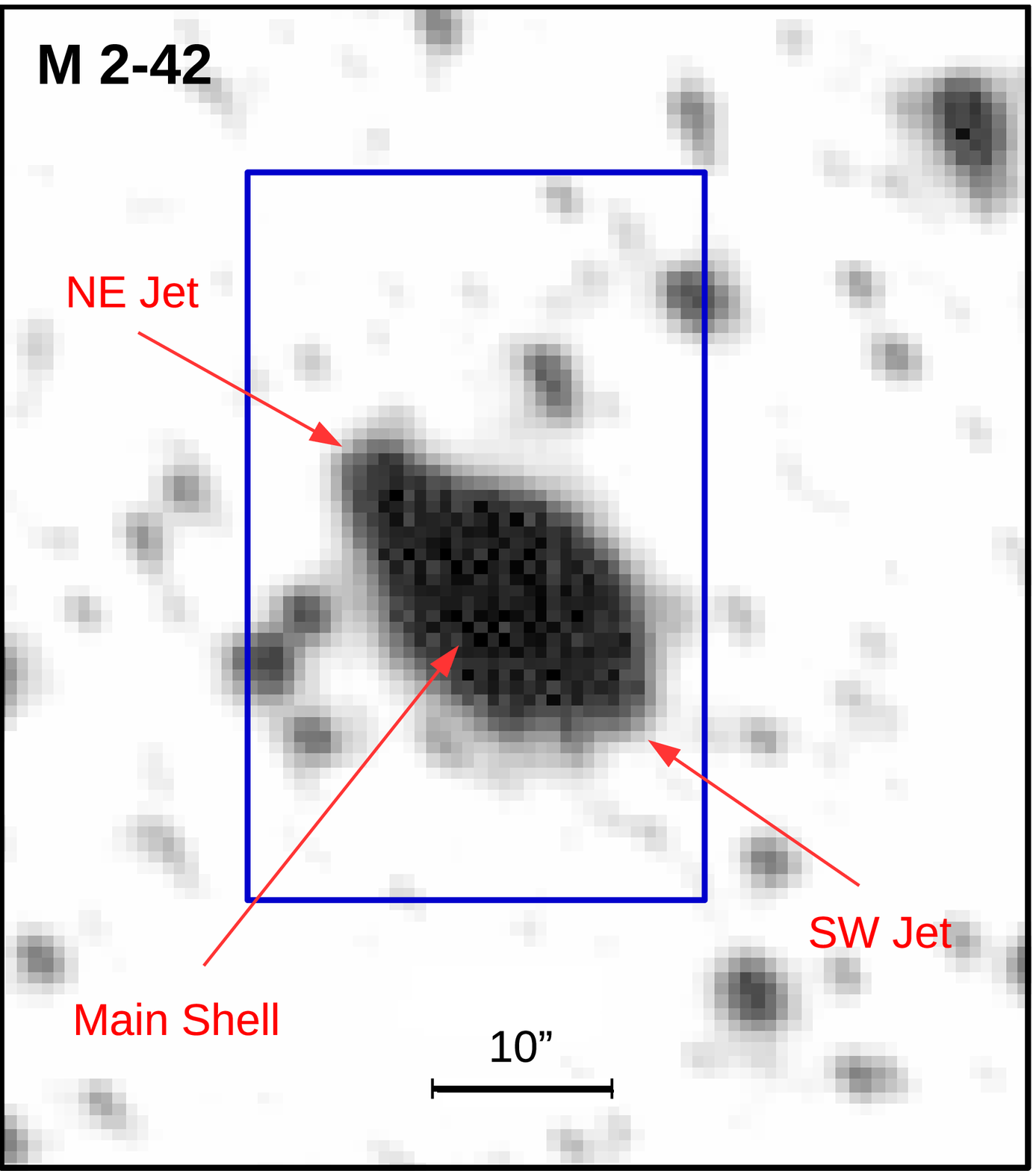}\\
\includegraphics[width=2.4in]{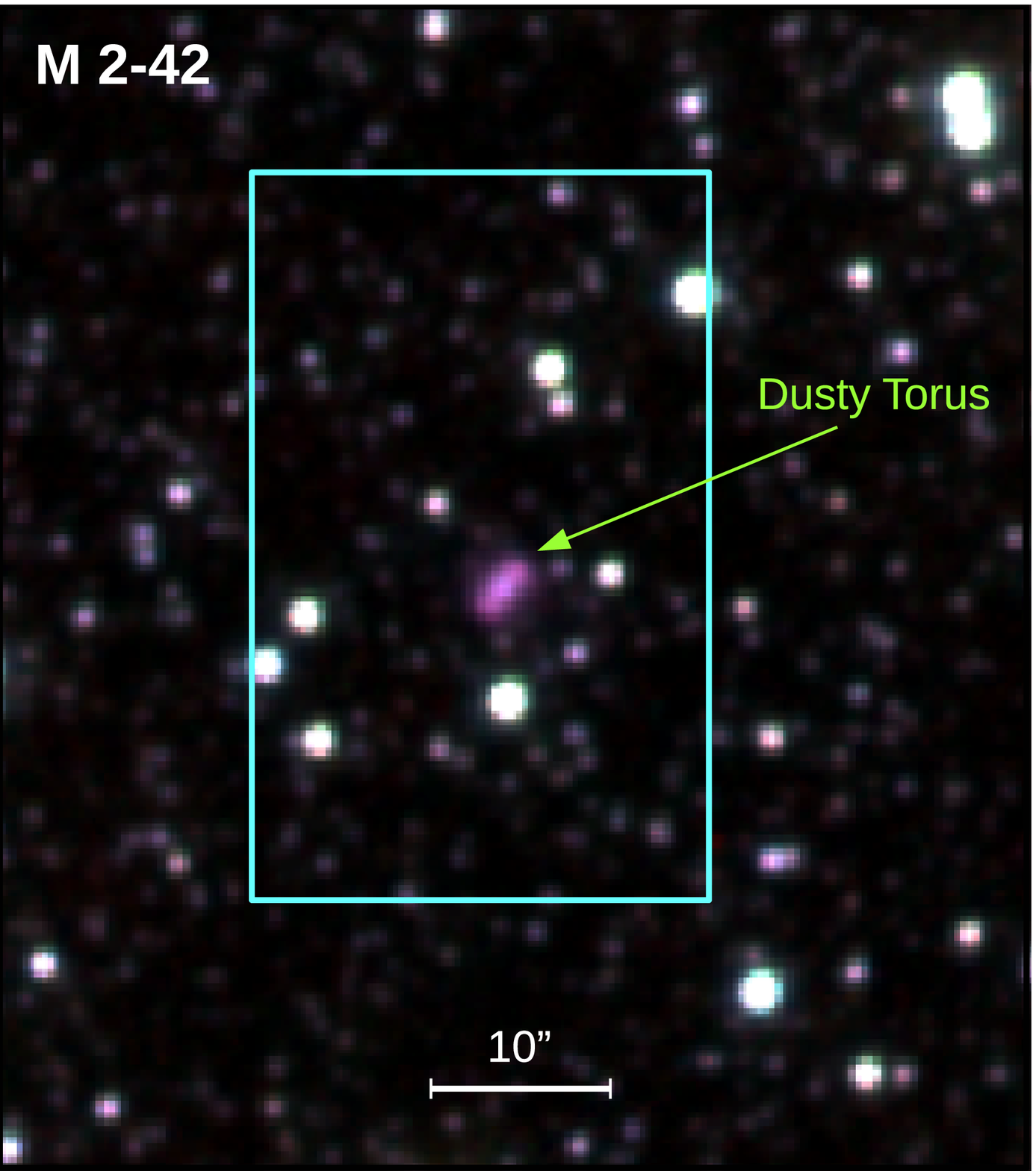}
\caption{Top panel: the H$\alpha$ image obtained from the SHS \citep{Parker2005} with the morphological features labeled. The rectangle shows the $25\arcsec \times 38\arcsec$ WiFeS field of view observed using the ANU 2.3-m telescope in 2010 April. The image scale is shown by a solid line. North is up and east is toward the left-hand side. Bottom panel: the \textit{JHK}$_{\rm s}$ image obtained from the VVV Survey \citep{Saito2012} with the compact dusty torus labeled. The red, green, and blue colors are assigned to the \textit{K}$_{\rm s}$, \textit{H}, and \textit{J}, respectively.
\label{fig:m2_42:shs}%
}%
\end{center}
\end{figure}

M2-42 (\,=\,PN\,G008.2$-$04.8 = Hen\,2-393 = VV\,177 = Sa\,2-331) was discovered as a planetary nebula (PN) by \citet{Minkowski1947}. The H$\alpha$ image, Fig. \ref{fig:m2_42:shs} (top panel), obtained from the AAO/UKST SuperCOSMOS H$\alpha$ Sky Survey \citep[SHS;][]{Parker2005} revealed an elliptical morphological structure with a clear extension to the north east, suggesting the presence of bipolar outflows. The long-slit data from the San Pedro M\'{a}rtir kinematic catalog \citep[SPM;][]{Lopez2012a} disclosed the presence of a dense torus-like component and collimated bipolar outflows \citep{Akras2012}. The \textit{JHK}$_{\rm s}$ image, Fig. \ref{fig:m2_42:shs} (bottom panel), obtained from the VISTA variables in the V\'{i}a L\'{a}ctea Survey \citep[VVV;][]{Saito2012} also shows the presence of a compact dusty torus embedded in the main shell.

\citet{Wang2007} carried out plasma diagnostics and abundance analysis of M2-42 using deep long-slit optical spectroscopy. They derived a mean electron density of $N_{\rm e}\simeq3 \times 10^3$\,cm$^{-3}$, and an electron temperature of $T_{\rm e}$\,=\,9\,350 K from the [N\,{\sc ii}] line ratio, which is consistent with those of other PNe \citep[see e.g.][]{Kingsburgh1994}. The oxygen abundance of O/H~=~$5.62 \times 10^{-4}$ derived by \citet{Wang2007} is slightly above the solar metallicity, while N/O\,=\,0.32 corresponds to a non-Type I PN \citep[based on N/O $< 0.8$;][]{Kingsburgh1994}.

The central star of M2-42 depicts weak emission-line star characteristics \citep[\textit{wels} defined by][]{Tylenda1993} dominated by nitrogen and helium \citep{Depew2011}. The nebular spectrum of moderate excitation, $I$(5007$)=807$ on a scale where $I$(H$\beta)=100$ \citep{Wang2007}, is related to an excitation class of 3.6 \citep{Dopita1990}, and a stellar temperature of 74\,kK \citep{Dopita1991} or 69\,kK \citep{Reid2010}. 
Based on the Energy-Balance method, \citet{Preite-Martinez1989} estimated a stellar temperature of 74.9 kK. According to \citet{Tylenda1991}, the central star has a B magnitude of 18.2. Using the H\,{\sc i} Zanstra method, \citet{Tylenda1991a} derived  a stellar temperature of 56 kK and a luminosity of $\log L/$L$_{\bigodot} = 2.87$, which correspond to a current core mass of 0.62M$_{\bigodot}$. 

Based on its angular diameter and radio brightness (6 cm), \citet{Acker1991a} suggested that M2-42 is most likely located in the Galactic bulge. \citet{Cahn1992} estimated a distance of 8754\,pc to the PN, which places it near to the Galactic center. The most recent distance estimation by \citet{Stanghellini2008} yielded a distance of 9444\,pc. Moreover, we estimate a distance of 7400$^{+570}_{-550}$ from the H$\alpha$ surface brightness-radius relation for a sample of 332 PNe \citep{Frew2016}, total flux value of $\log F($H$\alpha)= -11.39$ erg\,cm$^{-2}$\,s$^{-1}$ \citep{Frew2013}, $c$(\hb)$=0.99$ \citep{Wang2007}, and angular radius of 2 arcsec \citep{Stanghellini2008}. Therefore, it could be a Galactic Bulge PN (GBPN). 

In this paper, we present our integral field spectroscopy of M2-42, from which we determine ionization and kinematic properties of the nebula and its collimated outflows. In Section~\ref{m2_42:sec:observations}, we present the observations together with the physical and chemical conditions, stellar characteristics, and kinematic results derived from our data. Section~\ref{m2_42:sec:model} describes the morpho-kinematic model of M2-42 and, finally, in Section~\ref{m2_42:sec:discussion} we draw our conclusion. 

\newpage

\section{Observations}
\label{m2_42:sec:observations}

Moderate resolution, integral field observations were obtained on 2010 April 22 under program number 1100147 (PI: Q.\,A.~Parker) with the Wide Field Spectrograph \citep[WiFeS;][]{Dopita2007,Dopita2010} mounted on the Australian National University (ANU) 2.3 m telescope at Siding Spring Observatory. CCD chips with $4096 \times 4096$ pixels are used as detectors. The spectrograph samples $0\farcs5$ along each of twenty five $38\arcsec \times 1\arcsec$  slitlets, which provides a field of view of $25\arcsec \times 38\arcsec$ and a spatial resolution element of  $1\farcs0\times0\farcs5$. Each slitlet is designed to project to 2 pixels on the CCD chips, yielding a reconstructed point-spread function with a full width at half maximum (FWHM) of $\sim 2\arcsec$.

Figure \ref{fig:m2_42:shs} shows the WiFeS areal footprint used for our study. The main shell, the northeast (NE) jet and the southwest (SW) jet are also labeled on the figure. We used the spectral resolution  of $R\sim 7000$, covering  $\lambda\lambda$4415--5589\,{\AA} in the blue channel and $\lambda\lambda$5222--7070\,{\AA} in the red channel. The red spectrum has a linear wavelength dispersion per pixel of $0.45$ {\AA}, which yields a resolution of $\sim 20$ km\,s${}^{-1}$ in velocity channels. The exposure time of 20\,minutes used for our observation yields a signal-to-noise ratio of S/N\,$\gtrsim 10$ for the [N\,{\sc ii}] emission line. Data reduction was performed with the \textsf{wifes} \textsc{iraf} package \citep[described by][]{Danehkar2013,Danehkar2014a}.

\begin{figure}
\begin{center}
\includegraphics[width=1.70in]{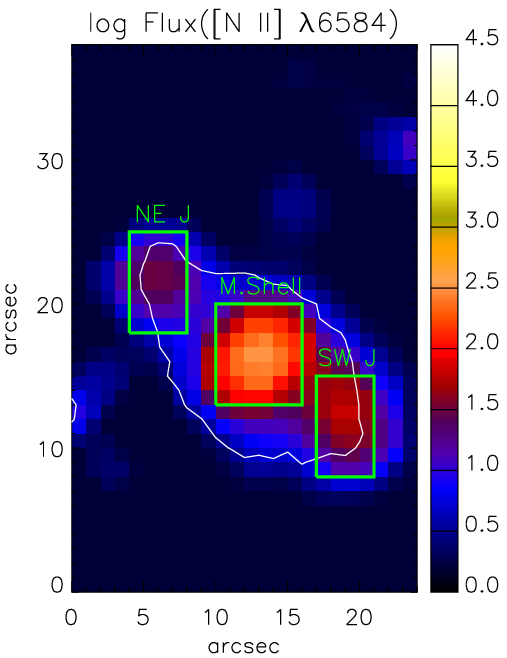}%
\includegraphics[width=1.70in]{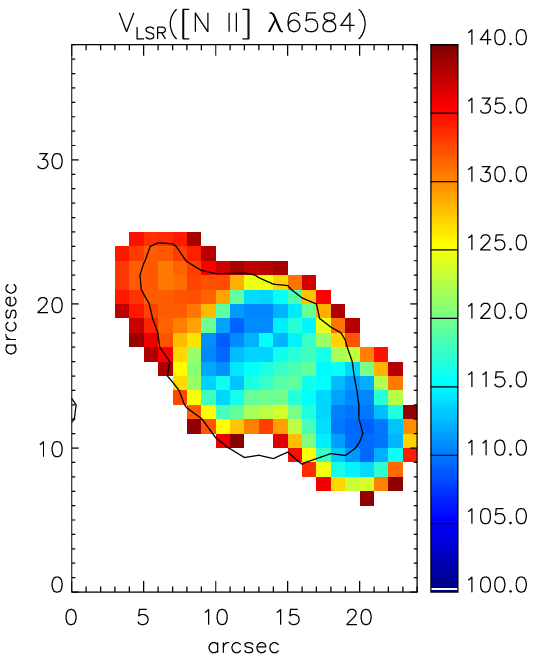}\\
\caption{From left to right, spatial distribution maps of flux intensity and LSR velocity of [N\,{\sc ii}] $\lambda$6584. Flux unit is in logarithm of $10^{-15}$~erg\,s${}^{-1}$\,cm${}^{-2}$\,spaxel${}^{-1}$ and velocity in km\,s${}^{-1}$. The rectangles show apertures used to extract fluxes across the main shell ($6\arcsec \times 7\arcsec$), and the NE and SW jets ($4\arcsec \times 7\arcsec$). The white/black contour lines show the distribution of the narrow-band emission of H$\alpha$ in arbitrary unit obtained from the SHS. North is up and east is toward the left-hand side. 
\label{fig:m2_42:ifu_map}%
}%
\end{center}
\end{figure}

\begin{table*}
\caption{\label{m2_42:tab:obs_lines}Observed line fluxes $F(\lambda)$ and dereddened fluxes $I(\lambda)$ measured from the apertures shown in Fig.\,\ref{fig:m2_42:ifu_map}.}
\centering
\begin{tabular}{lccccccccc}
\hline\hline
\noalign{\smallskip}
$\lambda_{0}$({\AA}) &ID	&Mult	&\multicolumn{2}{c}{Main Shell}&\multicolumn{2}{c}{NE Jet}&\multicolumn{2}{c}{SW Jet}\\
\noalign{\smallskip}
&	&	&$F(\lambda)$&$I(\lambda)$&$F(\lambda)$&$I(\lambda)$&$F(\lambda)$&$I(\lambda)$\\
\noalign{\smallskip}
\hline 
\noalign{\smallskip}
4471.50	&\hei	&V14	&4.56	&5.67	&4.07	&4.97	&3.75	&4.58\\
4609.44	&\oii	&V92a	&0.03	&0.04	&--	&--	&--	&--\\
4634.14	&\niii	&V2	&0.13	&0.15	&--	&--	&--	&--\\
4649.13	&\oii	&V1	&0.35	&0.39	&--	&--	&--	&--\\
4676.23	&\oii	&V1	&0.09	&0.10	&--	&--	&--	&--\\
4685.68	&\heii	&3.4	&0.30	&0.34	&0.34	&0.37	&--	&--\\
4740.17	&\fariv	&F1	&0.51	&0.54	&--	&--	&--	&--\\
4861.33	&\hi\,4-2	&H4	&100.00	&100.00	&100.00	&100.00	&100.00	&100.00\\
4881.11	&\ffeiii	&F2	&0.05	&0.05	&--	&--	&--	&--\\
4921.93	&\hei	&V48	&1.45	&1.41	&--	&--	&1.68	&1.63\\
4958.91	&\foiii	&F1	&226.94	&214.89	&172.83	&164.37	&193.45	&184.01\\
5006.84	&\foiii	&F1	&728.99	&672.08	&529.92	&491.74	&589.62	&547.27\\
5666.63	&\nii	&V3	&0.07	&0.04	&--	&--	&--	&--\\
5679.56	&\nii	&V3	&0.11	&0.07	&--	&--	&--	&--\\
5754.60	&\fnii	&F3	&1.03	&0.66	&2.21	&1.47	&1.35	&0.90\\
5875.66	&\hei	&V11	&24.78	&15.20	&23.82	&15.20	&25.96	&16.58\\
6101.83	&\fkiv	&F1	&0.05	&0.03	&--	&--	&--	&--\\
6312.10	&\fsiii	&F3	&2.24	&1.18	&--	&--	&--	&--\\
6461.95	&\cii	&V17.04	&0.12	&0.06	&--	&--	&--	&--\\
6548.10	&\fnii	&F1	&26.55	&12.87	&82.73	&42.49	&56.74	&29.21\\
6562.77	&\hi\,3-2  	&H3	&--	&286.00	&564.29	&288.58	&564.65	&289.38\\
6583.50	&\fnii	&F1	&81.59	&39.10	&258.01	&131.14	&174.03	&88.64\\
6678.16	&\hei	&V46	&8.60	&4.00	&8.59	&4.25	&8.81	&4.36\\
6716.44	&\fsii	&F2	&8.16	&3.75	&39.09	&19.11	&28.82	&14.12\\
6730.82	&\fsii	&F2	&12.97	&5.93	&36.51	&17.78	&30.02	&14.65\\
\noalign{\smallskip}
\multicolumn{3}{l}{$c$(\hb)}&	&0.989	&	&0.910	&	&0.907\\
\noalign{\smallskip}
\hline
\end{tabular}
\end{table*}

Table~\ref{m2_42:tab:obs_lines} presents a full list of observed line fluxes measured from three different apertures shown in Fig.\,\ref{fig:m2_42:ifu_map}: the main shell ($6\arcsec \times 7\arcsec$), the NE jet ($4\arcsec \times 7\arcsec$) and the SW jet ($4\arcsec \times 7\arcsec$). The laboratory wavelength, emission line identification and multiplet number are given in columns 1--3, respectively. Columns 4--9 present the observed line fluxes $F(\lambda)$ and the dereddened fluxes $I(\lambda)$ after correction for interstellar extinction for the three different regions, respectively. All fluxes are given relative to H$\beta$, on a scale where ${\rm H}\beta=100$. To extract the observed line fluxes, we applied a single Gaussian profile to each line. The logarithmic extinction $c({\rm H}\beta)$ was calculated from the Balmer flux ratio H$\alpha$/H$\beta$. However, we adopted the extinction $c$(\hb)$=0.989$ derived by \citet{Wang2007} for the main shell since the H$\alpha$ emission line was saturated over the main shell area.

\subsection{Physical and chemical conditions}

Electron temperature $T_{\rm e}$ and electron density $N_{\rm e}$ for the different regions of M2-42 are presented in Table~\ref{m2_42:tab:te_ne_abund}. The electron temperatures and densities were obtained using the \textsc{equib} code \citep{Howarth1981} from the \fnii\ nebular to auroral line ratio and the \fsii\ doublet line ratio, respectively. The electron temperature \elt(\fnii)$_{\rm corr}$  was corrected for recombination contribution to the auroral line using the formula given by \citet{Liu2000} and the ionic abundance ${\rm N^{++}}/{\rm H^{+}}$ derived from the N\,{\sc ii} lines. The values of $N_{\rm e}($\fsii$)=3150$\,cm$^{-3}$ and $T_{\rm e}($\fnii$)_{\rm corr}=9600$\,K are in agreement with $N_{\rm e}($\fsii$)= 3240$\,cm$^{-3}$ and $T_{\rm e}($\fnii$)=9350$\,K derived by \citet{Wang2007}. Additionally, we determined the physical conditions of the NE and SW jets. The jets show a mean electron temperature of $8840 \pm 180$\,K, which is 760\,K lower than that of the main shell, whereas their mean electron density of $595 \pm 125$\,cm$^{-3}$ is by a factor of five lower than that of the main shell. 

Table \ref{m2_42:tab:te_ne_abund} also lists the ionic abundances X${}^{i+}$/H${}^{+}$ derived from collisionally excited lines (CELs) and optical recombination lines (ORLs). We used the \textsc{equib} code to calculate the ionic abundances. We adopted the physical conditions, $T_{\rm e}$ (\elt\,$_{\rm corr}$ for the main shell) and $N_{\rm e}$, derived from CELs. The atomic data sets used for plasma diagnostics and abundances analysis are the same as those used by \citet[][Chapter 3]{Danehkar2014b}. 

Our value of He$^{+}$/H$^{+}=0.105$ for the main shell is in good agreement with He$^{+}$/H$^{+}$~=~0.107 derived by \citet{Wang2007}. However, they derived O$^{++}$/H$^{+}$~=~$5.27 \times 10^{-4}$, which is twice our value. This could be due to the different atomic data used by them. Our values of N$^{+}$/H$^{+}$, S$^{+}$/H$^{+}$ and Ar$^{3+}$/H$^{+}$ are in reasonable agreement with N$^{+}$/H$^{+}$~=~$1.03 \times 10^{-4}$, S$^{+}$/H$^{+}$~=~$4.96\times 10^{-7}$ and Ar$^{3+}$/H$^{+}$~=~$1.59 \times 10^{-7}$ obtained by \citet{Wang2007}. Note that a slit with a width of $2\arcsec$ used by \citet{Wang2007} is not completely related to the main shell. We see that the abundance discrepancy factor for O$^{++}$, ${\rm ADF}({\rm O}^{++}) \equiv ({\rm O}^{++}/{\rm H}^{+})_{\rm ORL} / ({\rm O}^{++}/{\rm H}^{+})_{\rm CEL}= 3.14$, is in agreement with ${\rm ADF}({\rm O}^{++})=2.09$ \citep{Wang2007}.
Moreover, our abundance ratio of (N$^{++}$/O$^{++})_{\rm ORL}=0.388$ derived from ORLs is in excellent agreement with (N$^{++}$/O$^{++})_{\rm ORL}=0.399$ obtained by \citet{Wang2007}. Although He$^{+}$/H$^{+}$ and O$^{++}$/H$^{+}$ derived from the jets are similar to those of the main shell, N$^{+}$/H$^{+}$ and S$^{+}$/H$^{+}$ derived from the jets are about three times higher than those of the main shell. These ionization features of the bipolar collimated jets are typical of fast, low-ionization emission regions \citep[FLIERs;][]{Balick1993,Balick1994,Balick1998}. 

\begin{table}
\caption{\label{m2_42:tab:te_ne_abund}Electron temperature \elt, electron density \eld\ and ionic abundances derived from the dereddened fluxes listed in Table \ref{m2_42:tab:obs_lines}.}
\centering
\begin{tabular}{lccc}
\hline\hline
\noalign{\smallskip}
Parameter &{Main Shell}&{NE Jet}&{SW Jet}\\
\noalign{\smallskip}
\hline 
\noalign{\smallskip}
{\elt(\fnii)(K)}&10270	&9020	&8660\\
\noalign{\smallskip}
{\elt(\fnii)$_{\rm corr}$(K)}&9600	&--	&--\\
\noalign{\smallskip}
{\eld(\fsii)(cm$^{-3}$)}&3150	&470  &720\\
\noalign{\smallskip}
{(He$^{+}$/H$^{+}$)$_{\rm ORL}$}&0.105	&0.107	&0.110\\
\noalign{\smallskip}
{(N$^{+}$/H$^{+}$)$_{\rm CEL} \times 10^{5}$}&0.764&2.912	&2.236\\
\noalign{\smallskip}
{(O$^{++}$/H$^{+}$)$_{\rm CEL} \times 10^{4}$}&2.606	&2.469&3.208\\
\noalign{\smallskip}
{(S$^{+}$/H$^{+}$)$_{\rm CEL} \times 10^{6}$}&0.347	&1.150	&1.040\\
\noalign{\smallskip}
{(S$^{++}$/H$^{+}$)$_{\rm CEL} \times 10^{6}$}&3.116	&--	&--\\
\noalign{\smallskip}
{(Ar$^{3+}$/H$^{+}$)$_{\rm CEL} \times 10^{7}$}&1.871&--	&--\\
\noalign{\smallskip}
{(N$^{++}$/H$^{+}$)$_{\rm ORL} \times 10^{4}$}&3.175	&--	&--\\
\noalign{\smallskip}
{(O$^{++}$/H$^{+}$)$_{\rm ORL} \times 10^{4}$}&8.185	&--	&--\\
\noalign{\smallskip}
\hline
\end{tabular}
\vspace{15pt}
\end{table}

\subsection{Comments on stellar characteristics}

The stellar emission-line fluxes presented in Table \ref{m242:tab:cspn}, are measured from a spectrum integrated over an aperture ($3\arcsec \times 3\arcsec$) covering the central star in the WiFeS field. The emission line identification, wavelength, dereddened flux corrected for reddening using $c$(\hb)$=0.99$, equivalent width $W_{\lambda}$ ({\AA}), and FWHM ({\AA}) are given in columns 1--5, respectively. All fluxes are given relative to C\,{\sc iv} 5805, on a scale where C\,{\sc iv}\,5805\,=\,100. We note that the width of C\,{\sc iv} $\lambda$5805 is narrower than typical Wolf--Rayet central starts of PNe with the same stellar temperature \citep[see e.g.,][]{Crowther1998,Acker2003}, so it could be a \textit{wels} as identified by \citet{Depew2011}. Following the method used by \citet{Acker2003}, a terminal wind velocity of 640\,km\,s$^{-1}$ is deduced from FWHM$($C\,{\sc iv}\,$\lambda$5805$)=10.27$\,{\AA}. However, we get a terminal velocity of 1560\,km\,s$^{-1}$ from FWHM$($C\,{\sc iv}\,$\lambda$5805$)=25$\,{\AA} reported by \citet{Depew2011}. Although the C\,{\sc iii} $\lambda$5696 line is not detected, the C\,{\sc iii}/{\sc iv} $\lambda$4650 is possibly identified. The He\,{\sc ii} $\lambda$4686 line is fairly strong, but it could have a nebular origin. We see the presence of strong N\,{\sc iii} $\lambda\lambda$4634,4641 lines and weak N\,{\sc v} $\lambda\lambda$4603,4932 lines. Assuming that the He\,{\sc ii} $\lambda$4686 line is a stellar emission line, M2-42 could have a stellar characteristics similar to WN8-type stars of \citet{vanderHucht2001} based on $I($N\,{\sc iii}\,$\lambda$4641$)\gtrsim I($He\,{\sc ii}\,$\lambda$4686) and $I($N\,{\sc iii}$)\gg I($N\,{\sc v}$)$. However, N\,{\sc ii} $\lambda$3995 and N\,{\sc iv} $\lambda\lambda$3479--3484, 4058 lines are not in the wavelength coverage of our WiFeS observations, so we cannot certainly classify it as one of nitrogen sequences of Wolf--Rayet central stars of PNe.

\begin{table}
\caption{\label{m242:tab:cspn}Stellar emission-line fluxes $I(\lambda)$ on a scale where C\,{\sc iv}\,5805\,=\,100, equivalent width $W_{\lambda}$({\AA}), and FWHM ({\AA}). }
\centering
\begin{tabular}{lcrrc}
\hline\hline
\noalign{\smallskip}
Line        &$\lambda$({\AA})&  $I(\lambda)$& $W_{\lambda}$({\AA}) &  {FWHM({\AA})} \\ 
\noalign{\smallskip}
\hline
\noalign{\smallskip}
N\,{\sc v}	&4603	&$33.58$:	&$-1.16$	&$3.14$\\
N\,{\sc iii}	&4634	&$77.56$	&$-3.15$	&$1.25$\\
N\,{\sc iii}	&4641	&$283.41$	&$-10.91$	&$3.08$\\
C\,{\sc iii}/{\sc iv}	&4650	&$220.38$	&$-9.49$	&$2.97$\\
C\,{\sc iii}	&4655	&$42.05$	&$-1.94$	&$3.55$\\
C\,{\sc iv}	&4659	&$63.19$	&$-2.81$	&$2.84$\\
He\,{\sc ii}	&4686	&$139.46$	&$-8.17$	&$1.23$\\
N\,{\sc v}	&4932	&$57.20$:	&$-2.93$	&$2.47$\\
C\,{\sc iv}	&5805	&$100.00$	&$-5.62$	&$10.27$\\
C\,{\sc ii}	&6462	&$19.28$	&$-1.11$	&$2.09$\\
C\,{\sc iii}	&7037	&$16.29$:	&$-1.04$	&$5.01$\\
\noalign{\smallskip}
\hline
\end{tabular}
\vspace{15pt}
\end{table}

\subsection{Kinematic results}

We derived an expansion velocity of $V_{\rm HWHM}=20.2\pm1.3$ km\,s$^{-1}$ from the half width at half maximum (HWHM) for the [N\,{\sc ii}] $\lambda\lambda$6548,6584 and [S\,{\sc ii}] $\lambda\lambda$6716,6731 emission-line profiles integrated over the main shell ($6 \arcsec \times 7 \arcsec$). The local standard of rest (LSR) systemic velocity of the whole nebula was estimated to be at $122.9\pm 12$\,km\,s$^{-1}$, which is in agreement with $V_{\rm LSR}=133.1 \pm 13.3$\,km\,s$^{-1}$ measured by \citet{Durand1998}. The LSR velocity is defined as the line of sight radial velocity, transferred to the local standard of rest by correcting for the motions of the Earth and Sun. 

\begin{figure}
\begin{center}
\includegraphics[width=3.5in]{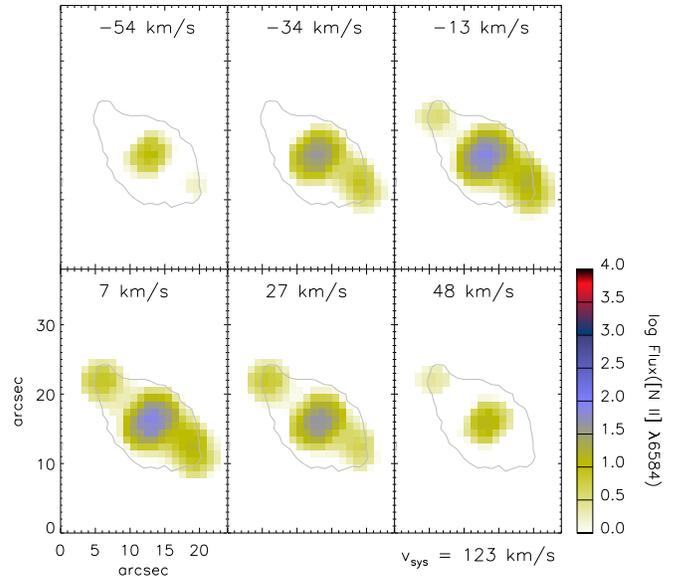}%
\caption{Velocity slices of M2-42 along the [N\,{\sc ii}] $\lambda$6584 emission-line profiles. The slices have a $\sim 20$ km\,s$^{-1}$ width, the central velocity is given at the top of each slice, and the LSR systemic velocity is $v_{\rm sys}=123$ km\,s$^{-1}$. The color bars show flux measurements in logarithm of $10^{-15}$~erg\,s${}^{-1}$\,cm${}^{-2}$\,spaxel${}^{-1}$. Velocity channels are in km\,s${}^{-1}$. The contours in the channel maps are the narrow-band H$\alpha$ emission in arbitrary unit obtained from the SHS.  North is up and east is toward the left-hand side.
\label{fig:m2_42:vmap}%
}%
\end{center}
\end{figure}

Figure~\ref{fig:m2_42:ifu_map} shows spatially resolved flux and velocity maps of M2-42 extracted from the [N\,{\sc ii}] $\lambda$6584 emission line across the WiFeS field. The observed radial velocity map was transferred to the LSR radial velocity. The white/black contour lines in the figures depict the 2D distribution of the H$\alpha$ emission obtained from the SHS, which can aid us in distinguishing the nebular border. As seen in Fig.\,\ref{fig:m2_42:ifu_map}, the kinematic map depicts an elliptical structure with a pair of collimated bipolar outflows, which is easily noticeable in the channel maps (see Fig. \ref{fig:m2_42:vmap}) and discussed below. 

Figure \ref{fig:m2_42:vmap} presents the flux intensity maps of the [N\,{\sc ii}] $\lambda$6584 emission lines on a logarithmic scale observed in a sequence of 6 velocity channels with a resolution of $\sim 20$ km\,s$^{-1}$, which can be used to identify different morphological components of the nebula. We subtracted the systemic velocity $v_{\rm sys}=123$ km\,s$^{-1}$ from the central velocity value given at the top of each channel. The stellar continuum map was also subtracted from the flux intensity maps. While there is a dense torus in the center, a pair of collimated bipolar outflows can be also identified in the velocity channels. The torus has a radius of $3\arcsec \pm 1 \arcsec$. This torus is clearly evident in the VVV \textit{J}, \textit{H}, and \textit{K}$_{\rm s}$ color combined image of M2-42 presented in Figure. \ref{fig:m2_42:shs} (bottom panel). We notice that the bipolar outflows are highly asymmetric, the SW jet apparently having a bow shock structure. While the NE jet reaches a distance of $12\arcsec \pm 2\arcsec$ from the nebular center, the distance of the SW jet from the central star is about 25\% shorter than the NE jet. Interaction with the interstellar medium (ISM) can lead to the formation of asymmetric bipolar outflows \citep[see e.g.][]{Wareing2007}. Both the jet components have similar brightness in the velocity channels. Brightness discontinuities are seen in the channels, where the bipolar outflows emerge from the main shell.

\section{Morpho-kinematic model}
\label{m2_42:sec:model}

We have used the morpho-kinematic modeling program \textsc{shape} (version 5.0) described in detail by \citet{Steffen2006} and \citet{Steffen2011}. This program has been used for modeling many PNe, such as NGC 2392 \citep{Garcia-Diaz2012}, NGC 3242 \citep{Gomez-Munoz2015}, Hen 2-113 and Hen 3-1333 \citep{Danehkar2015}. It uses interactively molded geometrical polygon meshes to generate three-dimensional structures of gaseous nebulae. The program produces several outputs that can be directly compared with observations, namely position-velocity diagrams, velocity channels and synthetic images. The modeling procedure consists of defining the geometry, assigning a density distribution and defining a velocity law. Geometrical and kinematic parameters are modified in a manual interactive process until a satisfactorily fitting model has been constructed.

\begin{figure}
\begin{center}
{\footnotesize (a) \textsc{shape} model}\\
\includegraphics[width=3.5in]{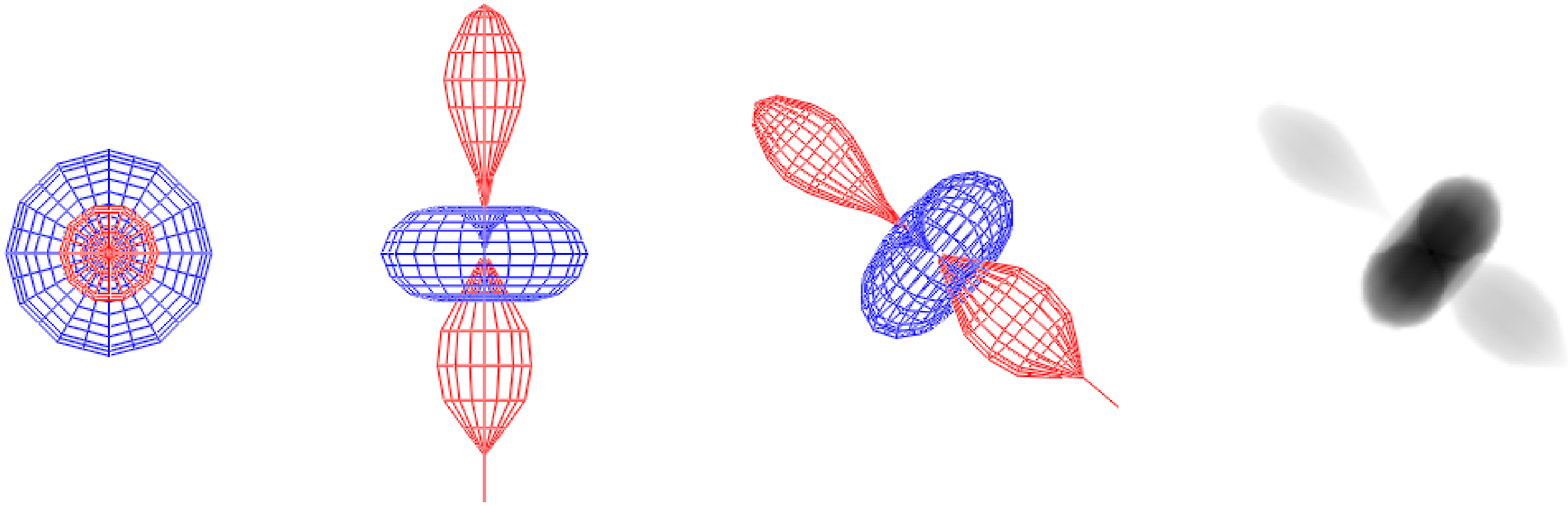}\\
{\footnotesize (b) Velocity channels}\\ 
\includegraphics[width=3.4in]{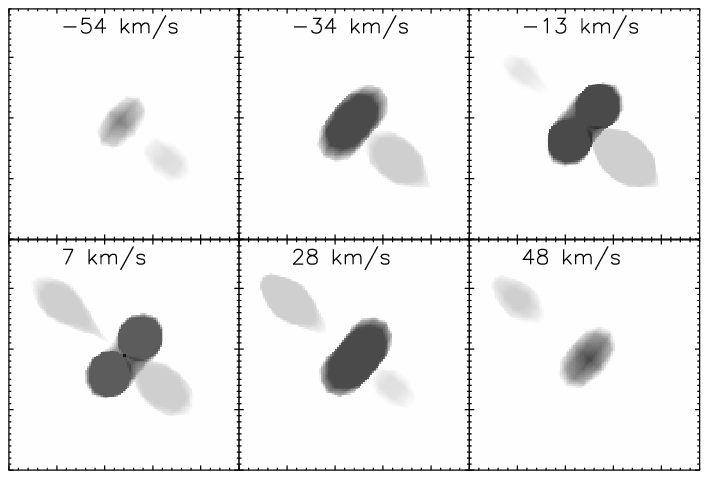}%
\caption{Top panels: \textsc{shape} mesh model of M2-42 before rendering at two different orientations (inclination: 0$^{\circ}$ and 90$^{\circ}$), the best-fitting inclination, and the corresponding rendered image, respectively.
Bottom panels: synthetic images at different velocity channels obtained from the best-fitting \textsc{shape} model. 
\label{m2_42:shape}%
}%
\end{center}
\end{figure}

Figure \ref{m2_42:shape} (a) shows the morpho-kinematic model before rendering at two different orientations (inclination: 0$^{\circ}$ and 90$^{\circ}$), and their best-fitting inclination, together with the result of the rendered model. The morpho-kinematic model consists of an equatorial dense torus (main shell) and a pair of asymmetric bipolar outflows. The values of the parameters of the final model are summarized in Table~\ref{m2_42:parameters}. For the velocity field, we assume a Hubble-type flow \citep{Steffen2009}. 

The velocity-channel maps of the final model are shown in Figure \ref{m2_42:shape} (b), where they can be directly compared with the observed velocity-resolved channel maps presented in Figure \ref{fig:m2_42:vmap}. The model maps are a good match  to the observational maps. The model successfully produces two kinematic components of the jets moving in opposite directions on both sides of the torus. From the morpho-kinematic model, we derived an inclination of $i=-82^{\circ}\pm 4^{\circ}$ with respect to the line of sight. Taking the inclination derived by the best-fitting model, we estimated a ``jet'' expansion velocity of $ 120 \pm 40$ km\,s$^{-1}$ with respect to the central star.

\begin{table}
\begin{center}
\caption{Parameters of the Morpho-kinematic model of M2-42.\label{m2_42:parameters}}
\begin{tabular}{lc}
\hline\hline
\noalign{\smallskip}
Parameter & Value \\
\noalign{\smallskip}
\hline
\noalign{\smallskip}
Inclination of major axis, $i$& $-82^{\circ} \pm 4^{\circ}$  \\
\noalign{\smallskip}
Position angle of major axis, PA				& $50^{\circ} \pm 5^{\circ}$  \\
\noalign{\smallskip}
Galactic position angle of major axis, GPA    & $112^{\circ}24\arcmin \pm 5^{\circ}$  \\
\noalign{\smallskip}
Outer radius of the main shell			& $3\pm1$ arcsec \\
\noalign{\smallskip}
NE Jet distance from the center				& $12\pm2$ arcsec \\
\noalign{\smallskip}
SW Jet distance from the center				& $9\pm2$ arcsec \\
\noalign{\smallskip}
Jet velocity from the center	  			& $120\pm40$ km\,s${}^{-1}$\\
\noalign{\smallskip}
\hline
\end{tabular}
\end{center}
\end{table}

As seen in Table~\ref{m2_42:parameters}, the symmetric axis of the bipolar outflows has a position angle (PA) of $50^{\circ} \pm 5^{\circ}$ measured from the north toward the east in the equatorial coordinate system (ECS). This leads to a Galactic position angle (GPA) of $112 \fdg 4$. The GPA is the position angle of the nebular symmetric axis projected on to the sky plane, measured from the North Galactic Pole toward the Galactic east. Note that ${\rm GPA}=90^{\circ}$ describes an alignment with the Galactic plane, whereas ${\rm GPA}=0^{\circ}$ is perpendicular to the Galactic plane. Therefore, the symmetric axis of M2-42 is roughly aligned with the Galactic plane. This alignment could have some implications for other studies of GBPNe \citep[see e.g.][]{Rees2013,Falceta-Gonccalves2014,Danehkar2016}.

\section{Summary and discussions}
\label{m2_42:sec:discussion}

In this paper, we present the spatially resolved observations of M2-42 obtained with the WiFeS on the ANU 2.3 m telescope. Using the velocity-resolved channel maps derived from the [N\,{\sc ii}] $\lambda$6584 emission line, a morpho-kinematic model has been developed which includes different morphological components of the nebula: a dense torus and a pair of asymmetric bipolar outflows in opposite directions. From the HWHM method, the torus is found to expand slowly at $20$\,km\,s$^{-1}$, almost in agreement with $15$\,km\,s$^{-1}$ derived by \citet{Akras2012}. From the reconstruction model, the trail of bipolar outflows was found to go along the direction of (GPA, $i$) $=$ ($112^{\circ}$, $-82^{\circ}$), which is very similar to the inclination of $i=77^{\circ}$ derived by \citet{Akras2012} based on the SPM long-slit data. We find a ``jet'' expansion velocity of $ 120 \pm 40$\,km\,s$^{-1}$ with respect to the nebular center, which is higher than the value of $70$\,km\,s$^{-1}$ estimated by \citet{Akras2012}. Moreover, we found that the SW jet, which moves toward us, has possibly a bow shock structure relating to the interaction with ISM \citep[see e.g.][]{Wareing2007}. 

An empirical analysis of the nebular spectra separately integrated over the three different regions shows that the mean density of the jets is a factor of five lower than that in the main shell. Although the abundances of singly ionized helium and doubly ionized oxygen are almost the same in both the shell and the jets, the singly ionized nitrogen and sulfur abundances derived from the jets are about three times higher than those obtained from the main shell. The similar ionization characteristics have been found in collimated jets emerged from other PNe \citep[see e.g.][]{Balick1993,Balick1994}. 

Nearly 10\% of Galactic PNe have been found to have the small-scale low-ionization structures in opposite directions on both sides of their central stars. Around  half of them are fast, highly collimated outflows with velocities of 30--200 km\,s$^{-1}$ relative to the main bodies, so called FLIERs \citep[][]{Balick1993,Balick1994,Balick1998}. Previously, \citet{Balick1994} claimed the presence of nitrogen enrichment by factors of 2--5 in the FLIERs of some PNe. However, \citet{Gonccalves2003} suggested that empirically derived nitrogen overabundance seen in FLIERs are a result of inaccurate ionization correction factors applied in the empirical analysis. \citet{Gonccalves2006} constructed a chemically homogeneous photoionization model of NGC\,7009, which can reproduce the ionization characteristics of its shell and FLIERs. Similarly, the enhancement of N$^{+}$/H$^{+}$ and S$^{+}$/H$^{+}$ in the FLIERs of M2-42 could be attributed to the geometry and density distribution rather than chemical inhomogeneities.

The previous observations of M2-42 showed that its central star is of \textit{wels} type \citep{Depew2011}. Moreover, we found that its stellar spectrum might be similar to the WN8 subclass of \citet{vanderHucht2001} based on $I($N\,{\sc iii}$)\gtrsim I($He\,{\sc ii}). The terminal wind velocity was also estimated to be about 640 \,km\,s$^{-1}$. However, our observations did not cover the N\,{\sc ii} and N\,{\sc iv} lines, which are necessary for the WN classification. This typical stellar characteristics and its point-symmetric morphology could be a result of a common-envelope evolutionary phase \citep[see e.g.][]{Nordhaus2006}. Currently, there is no evidence for binarity in M2-42. We believe that further observations of its central star will help develop a better stellar classification and also shed light on the mechanism producing its FLIERs.

\acknowledgments

A.D. acknowledges the award of a Research Excellence Scholarship from Macquarie University. Q.A.P. acknowledges support from Macquarie University and the Australian Astronomical Observatory (AAO). W.S. acknowledges support from grant UNAM-PAPIIT 101014. We would like to thank the staff at the ANU Siding Spring Observatory for their support. We acknowledge use of data from the VISTA telescope under ESO Survey programme ID 179.B-2002. We thank the anonymous referee whose suggestions and comments have greatly improved the paper.

\end{document}